\documentclass[twocolumn,final]{svjour3}         
\smartqed  
\usepackage{amstext}
\usepackage{graphicx}

\usepackage{caption}
\usepackage{subcaption}

\usepackage{natbib}

\usepackage{fix-cm}
\usepackage{mathptmx}      
\usepackage{color}

\usepackage{lineno}

\journalname{Microgravity Science and Technology}

\begin{document}


\title{Drop tower setup to study the diffusion-driven growth of a foam ball in supersaturated liquids in microgravity conditions}


\author{Patricia Vega-Mart\'{\i}nez $^{1}$ \and Javier Rodr\'{\i}guez-Rodr\'{\i}guez  $^{1}$ \and Devaraj van der Meer  $^{2}$ \and Matthias Sperl  $^{3}$}

\institute{\at $^{1}$ Fluid Mechanics Group, University Carlos III of Madrid, 28911 Legan\'{e}s, Madrid, Spain\\
 				\email{pvega@ing.uc3m.es} \\
$^{2}$ Physics of Fluids Group, MESA+ Research Institute, and J. M. Burgers Centre for Fluid Dynamics, University of Twente, P.O. Box 217, 7500 AE Enschede, The Netherlands\\
$^{3}$  Institut f\"ur Materialphysik im Weltraum, Deutsches Zentrum f\"ur Luft- und Raumfahrt, 51170 Cologne, Germany  \\
}

\date{Received: date / Accepted: date}

\maketitle


\begin{abstract}

The diffusion-driven growth of a foam ball is a phenomenon that appears in many manufacturing process as well as in a variety of geological phenomena.
Usually these processes are greatly affected by gravity, as foam is much lighter than the surrounding liquid.
However, the growth of the foam free of gravity effects is still very relevant, as it is connected to manufacturing in space and to the formation of rocks in meteorites and other small celestial bodies.
The aim of this research is to investigate experimentally the growth of a bubble cloud growing in a gas-supersaturated liquid in microgravity conditions.
Here, we describe the experiments carried out in the drop tower of the Center of Applied Space Technology and Microgravity (ZARM). In few words, a foam seed is formed with spark-induced cavitation in carbonated water, whose time evolution is recorded with two high-speed cameras.
Our preliminary results shed some light on how the size of the foam ball scales with time, in particular at times much longer than what could be studied in normal conditions, i.e. on the surface of the Earth, where the dynamics of the foam is already dominated by gravity after several milliseconds.
\end{abstract}


\section{Introduction}
\label{intro}

The diffusion-driven growth of a dense bubble cloud  immersed in a supersaturated gas-water solution is of interest to understand a variety of phenomena that range from industrial applications to geology. For instance, in photo- and electrocatalysis the physical mechanisms responsible of the growth of bubbles on the catalytic surface share many similarities with those driving the growth of a bubble cloud by pure diffusion. More fundamental applications are found in the field of planetary geology. For instance, to understand which mechanisms determine the amount of noble gases --in particular Helium-- found in meteorites and other small solar-system bodies, it is essential to study the diffusion of these gases outside the body through the complex multiphase flow ocurring during their solidification. Naturally, this diffusion process occurs in low-gravity conditions \citep{7}.

A bubble that forms part of a cloud and grows by diffusion in a supersaturated CO$_2$ solution increases its size at a pace slower than that of an isolated bubble, since it has to compete for the available CO$_2$ with its neighbors. Our previous experiments \citep{RodriguezRodriguez_etalPRL2014} suggest that, at short times, when the inter-bubble distance is relatively large, bubble sizes grow with the square root of the time, as predicted by the diffusion-driven regime \citep{3,4}. More interestingly, as the void fraction of the cloud grows, the growth rate departs from this regime in ways that are not fully understood \citep{5}, in contrast with the scaling found in the coarsening of dry foams at constant liquid fraction, where bubble sizes grow as $t^{1/2}$ at all times \citep{DurianWeitzPinePRA1991}. Under normal gravity conditions, this purely diffusive competition process is interrupted after a few milliseconds when the bubble cloud starts a buoyancy-induced rising motion, and advection --understood as the transport of dissolved gas by the fluid velocity field-- and mixing dominate thereafter the bubble-liquid gas transfer. In this paper, we describe a novel experimental setup aimed at exploring the diffusion-driven growth experimentally at times much longer than what is possible in normal conditions on Earth making use of a microgravity facility. The ultimate goal of this experiment is to obtain quantitative data that will serve us to validate theoretical models on the diffusion-driven growth of dense bubble clouds. It is worth pointing out that the void fractions explored here are smaller than those found in foams, which have been studied in the past in microgravity conditions both theoretically \citep{CoxVerbistMGST2003} and experimentally \citep{Saint-Jalmes_etalMGST2006}. Also the behavior of plateau borders, where the interfaces of adjacent bubbles meet, has been studied in the absence of gravity \citep{Barret_etalMGST2008}.

Although studying cavitation is not the main purpose of the experiment, we exploit this phenomenon to generate the bubble cloud. Cavitation in a microgravity environment was explored experimentally by \cite{Obreschkow_etalPRL2011}. In their experiments, they induced cavitation by focusing a laser pulse in the bulk liquid, whereas here we use a spark for that purpose. However, the main difference is that in the work by Obreschkow {\it et al.} the gas cavity disappears upon its collapse, whereas here the bubble fragments that result from the collapse become the nuclei from which the bubbles in the cloud will grow. This different behavior occurs because the water that we use is supersaturated, i.e. contains more CO$_2$ than what the liquid can dissolve, thus this gas fills the cavitation fragments and precludes their dissolution.

\begin{figure*}
\centering
\begin{subfigure}[b]{0.23\textwidth}
\includegraphics[width =\columnwidth,trim = 70 50 70 40,clip]{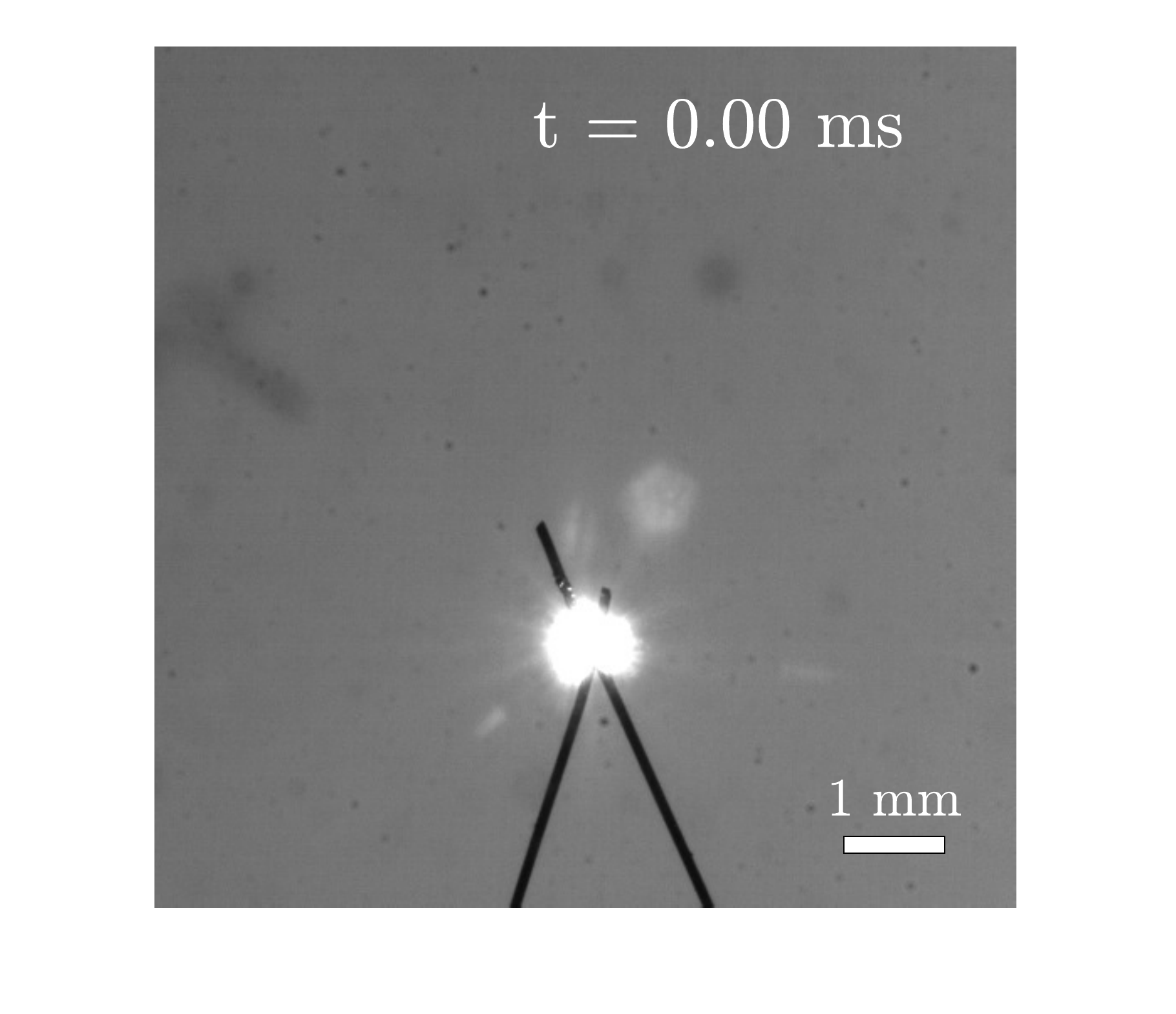}
\caption{\label{fig:1}}
\end{subfigure}
\begin{subfigure}[b]{0.23\textwidth}
\includegraphics[width =\textwidth,trim = 70 50 70 40,clip]{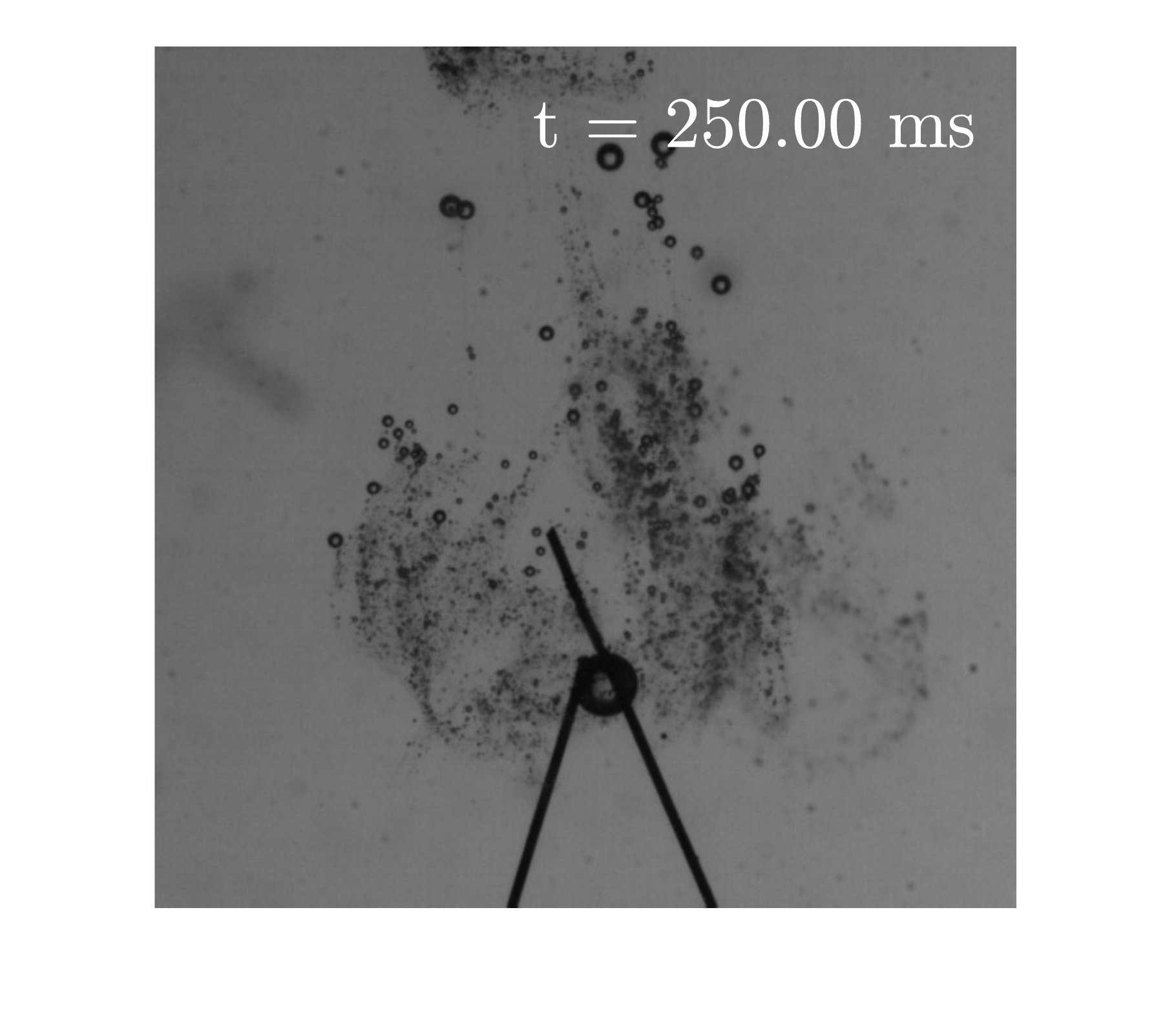}
\caption{\label{fig:2}}
\end{subfigure}
\begin{subfigure}[b]{0.23\textwidth}
\includegraphics[width =\textwidth,trim = 70 50 70 40,clip]{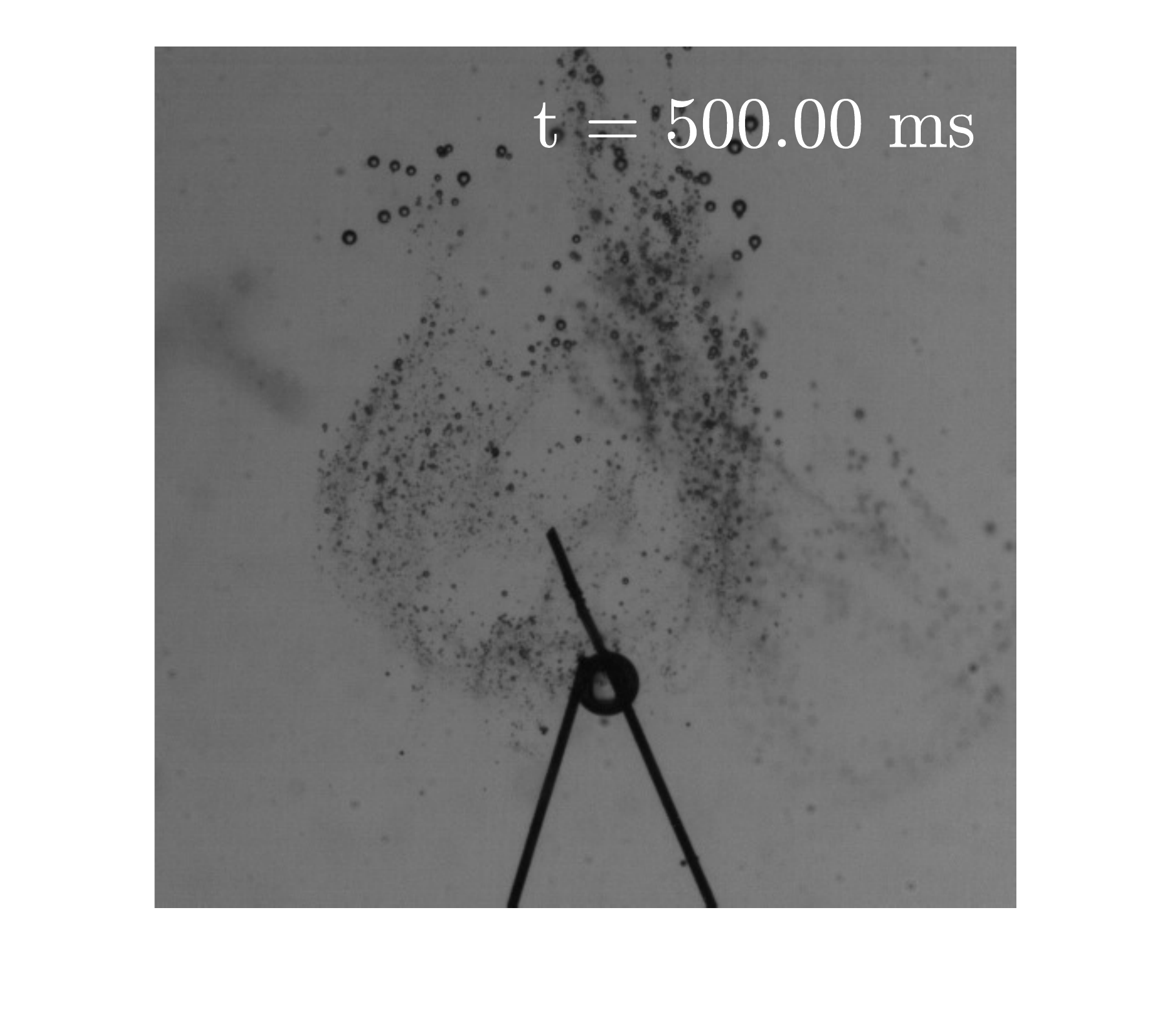}
\caption{\label{fig:3}}
\end{subfigure}
\begin{subfigure}[b]{0.23\textwidth}
\includegraphics[width =\textwidth,trim = 70 50 70 40,clip]{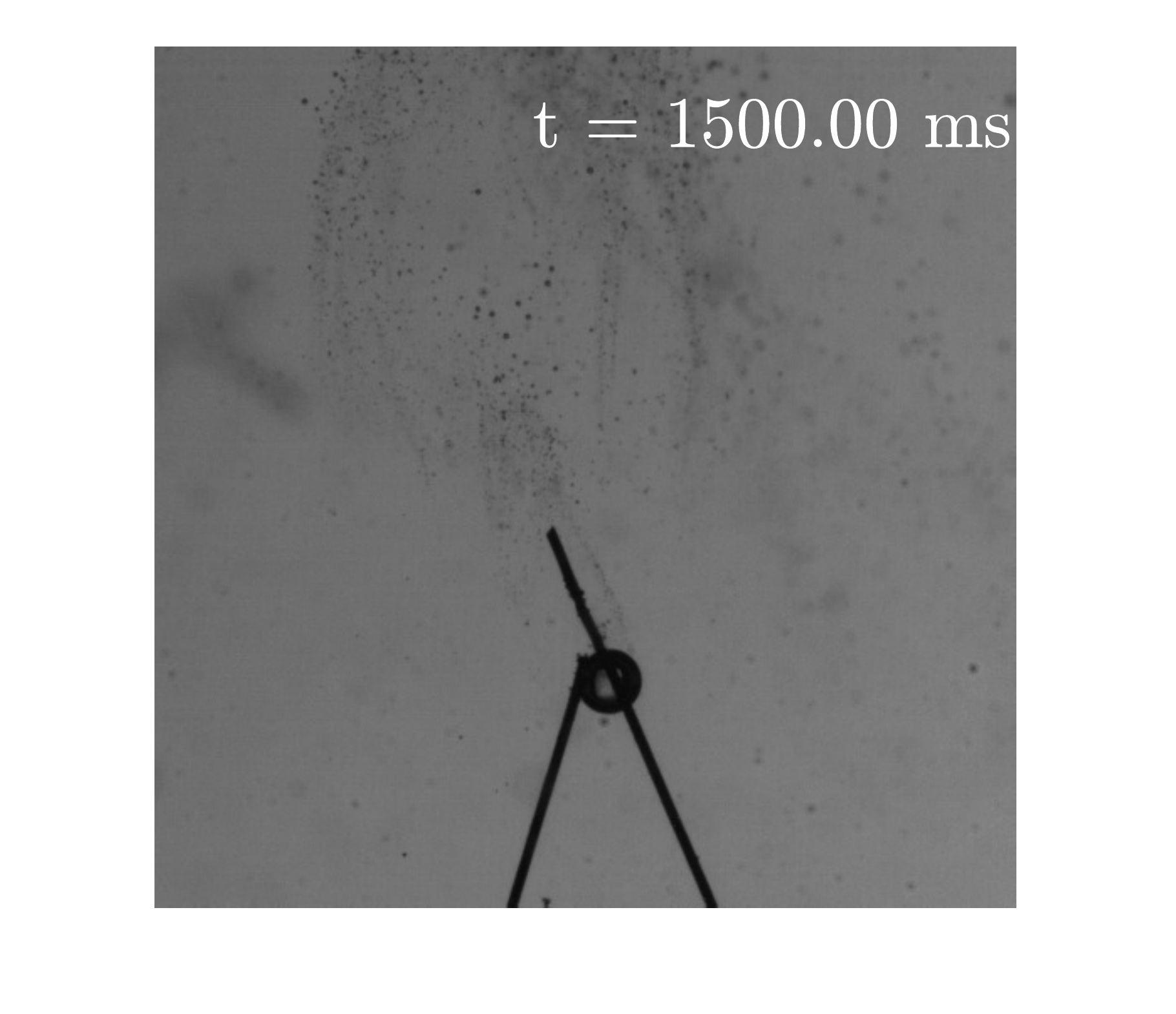}
\caption{\label{fig:4}}
\end{subfigure}

\begin{subfigure}[b]{0.23\textwidth}
\includegraphics[width =\textwidth,trim = 70 50 70 40,clip]{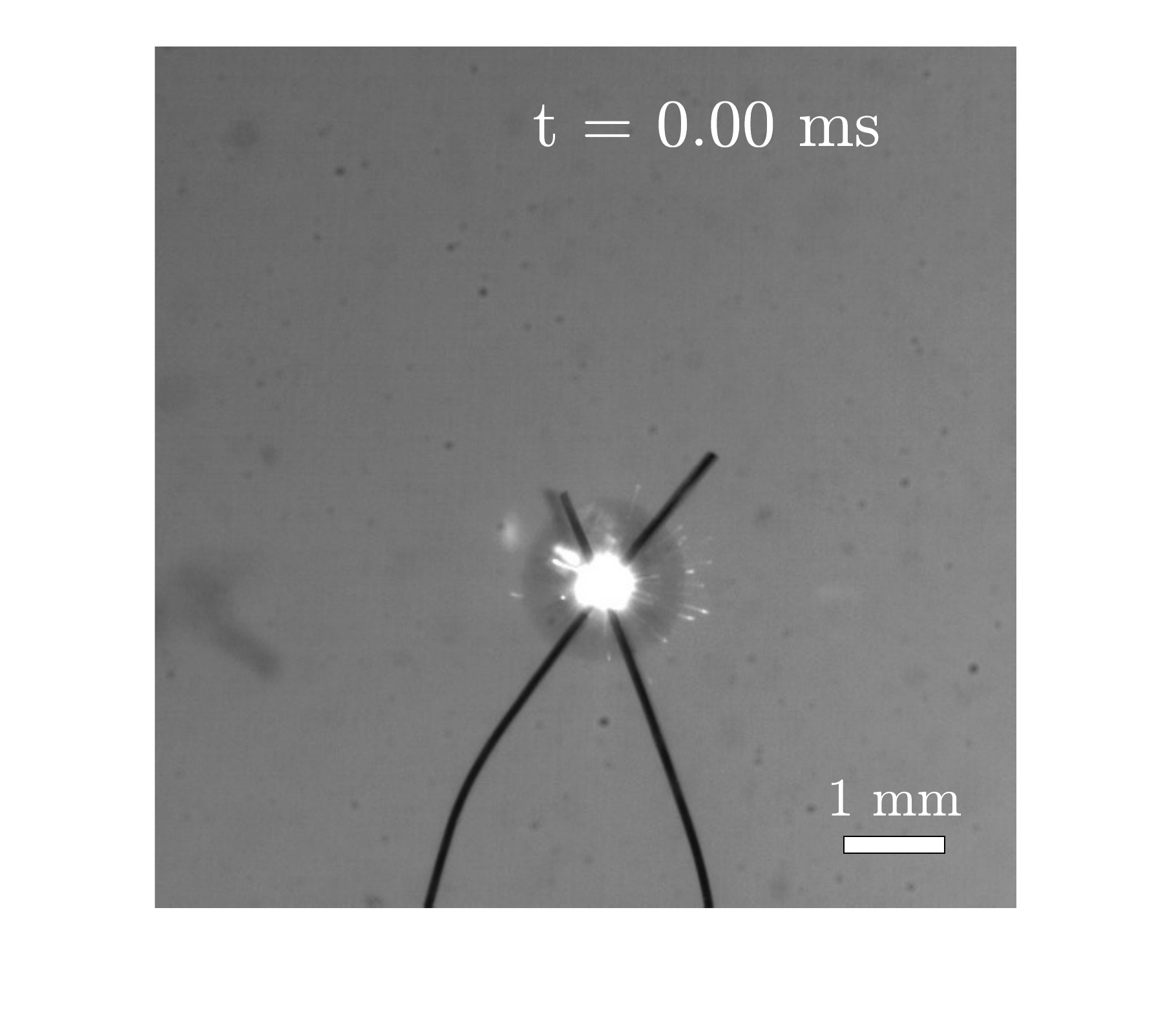}
\caption{\label{fig:5}}
\end{subfigure}
\begin{subfigure}[b]{0.23\textwidth}
\includegraphics[width =\textwidth,trim = 70 50 70 40,clip]{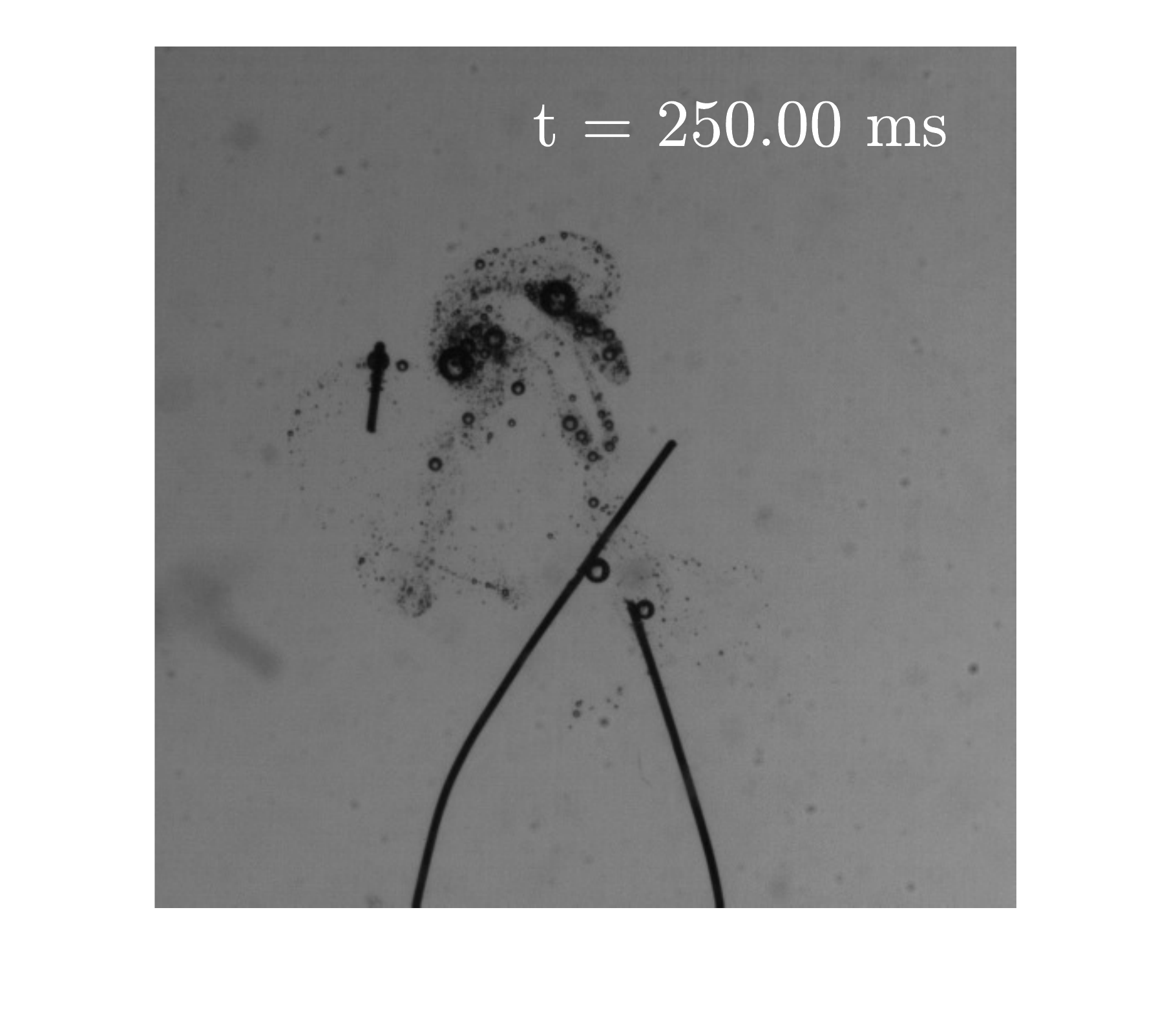}
\caption{\label{fig:6}}
\end{subfigure}
\begin{subfigure}[b]{0.23\textwidth}
\includegraphics[width =\textwidth,trim = 70 50 70 40,clip]{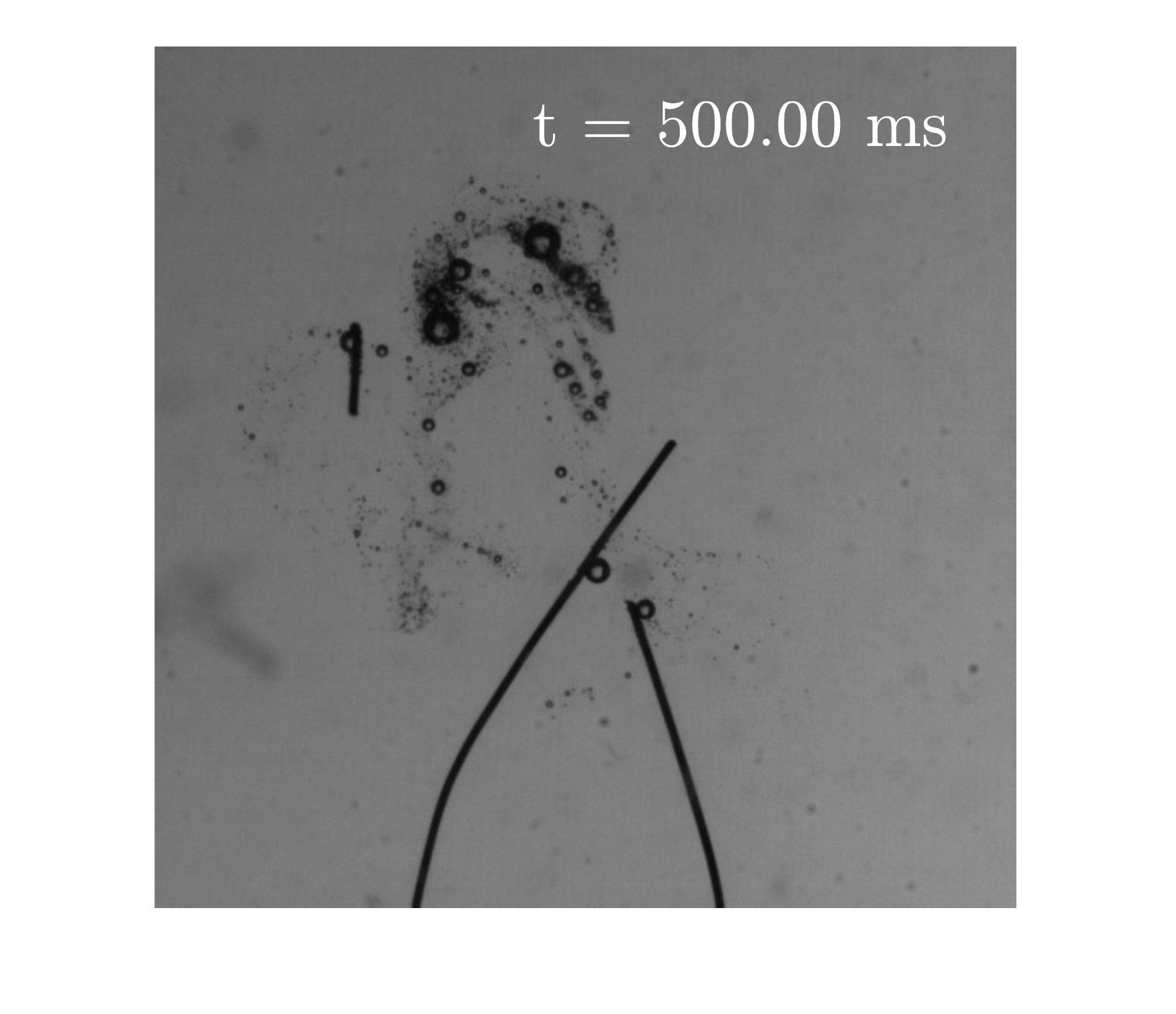}
\caption{\label{fig:7}}
\end{subfigure}
\begin{subfigure}[b]{0.23\textwidth}
\includegraphics[width =\textwidth,trim = 70 50 70 40,clip]{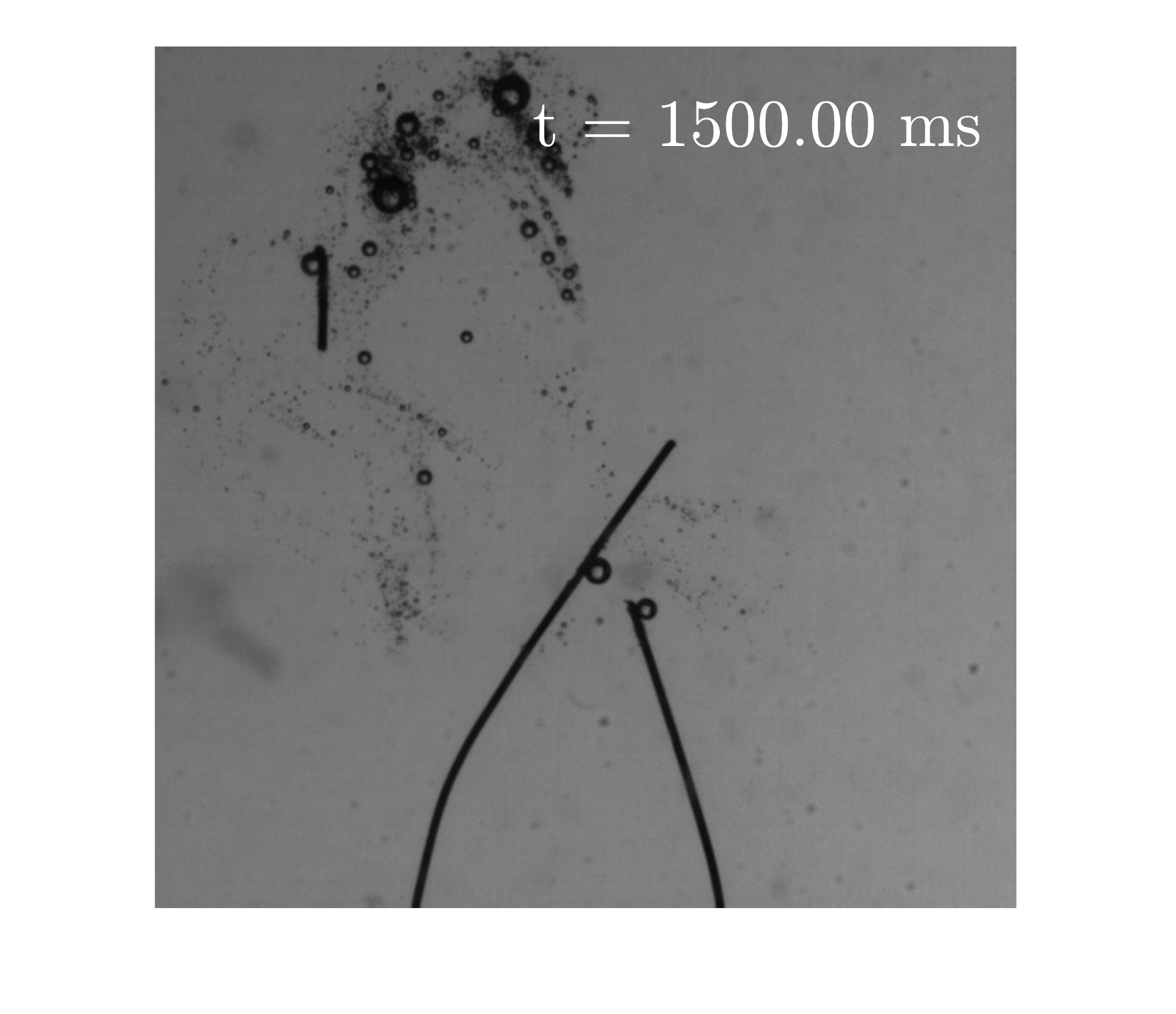}
\caption{\label{fig:8}}
\end{subfigure}
\caption{\label{fig:comparation} \textsl{Comparison between experiments in gravity [a - d] and microgravity [e - h] conditions.}}
\end{figure*}

\section{Setup of the experiment}
\label{setup}

\begin{figure}[ht!]
\centering
\includegraphics[width = 1\columnwidth]{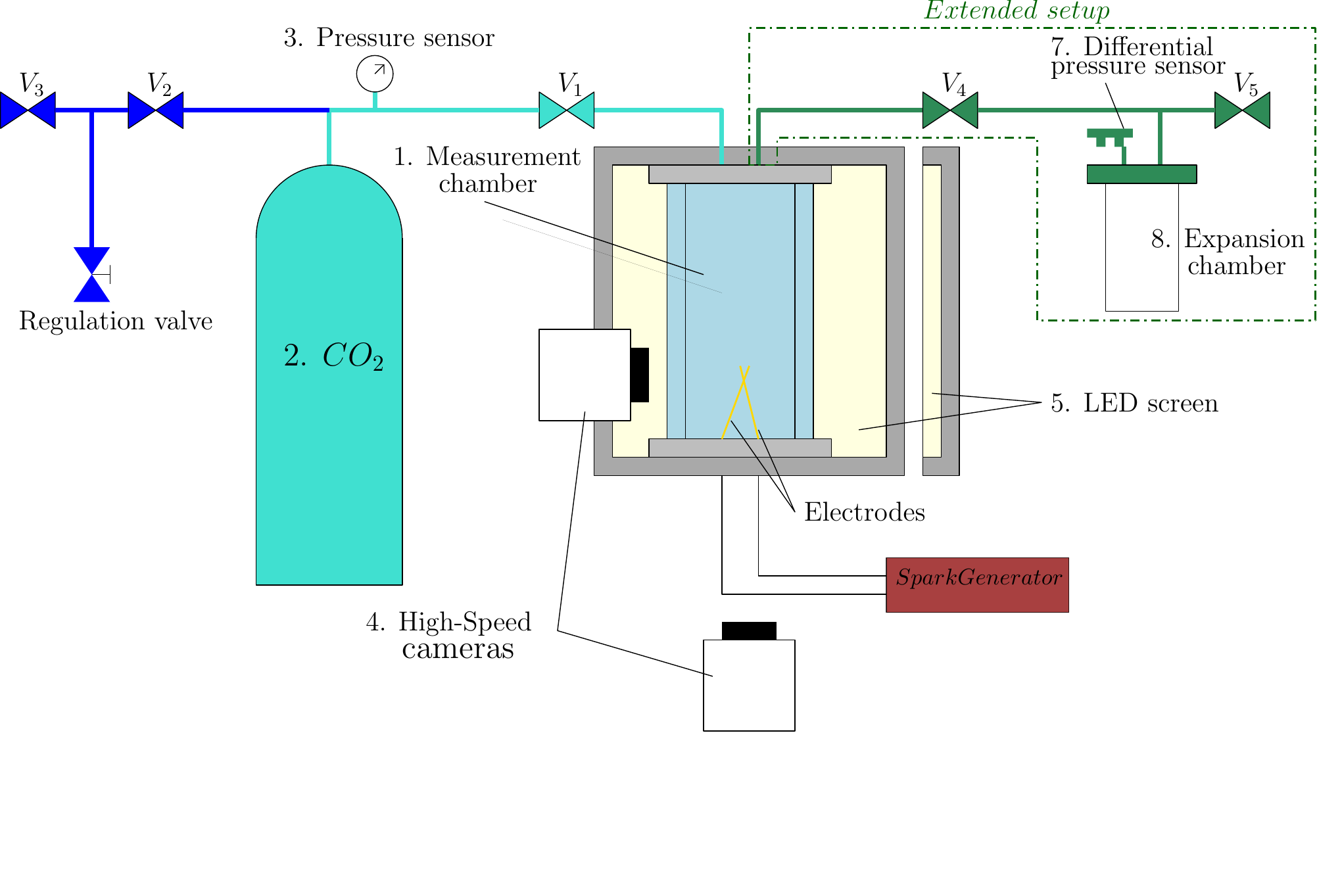}
\caption{\label{fig:sketch}\textsl{Layout of the experimental set-up. The measurement chamber is filled with carbonated water. At its bottom there are two electrodes connected to the spark generator device. The top of the chamber is connected to a pressurization and depressurization system. The two high-speed cameras shown in the sketch form an angle of 90$^{\,\rm o}$.}}
\end{figure}

\begin{figure}
\centering
\includegraphics[width = 0.9\columnwidth]{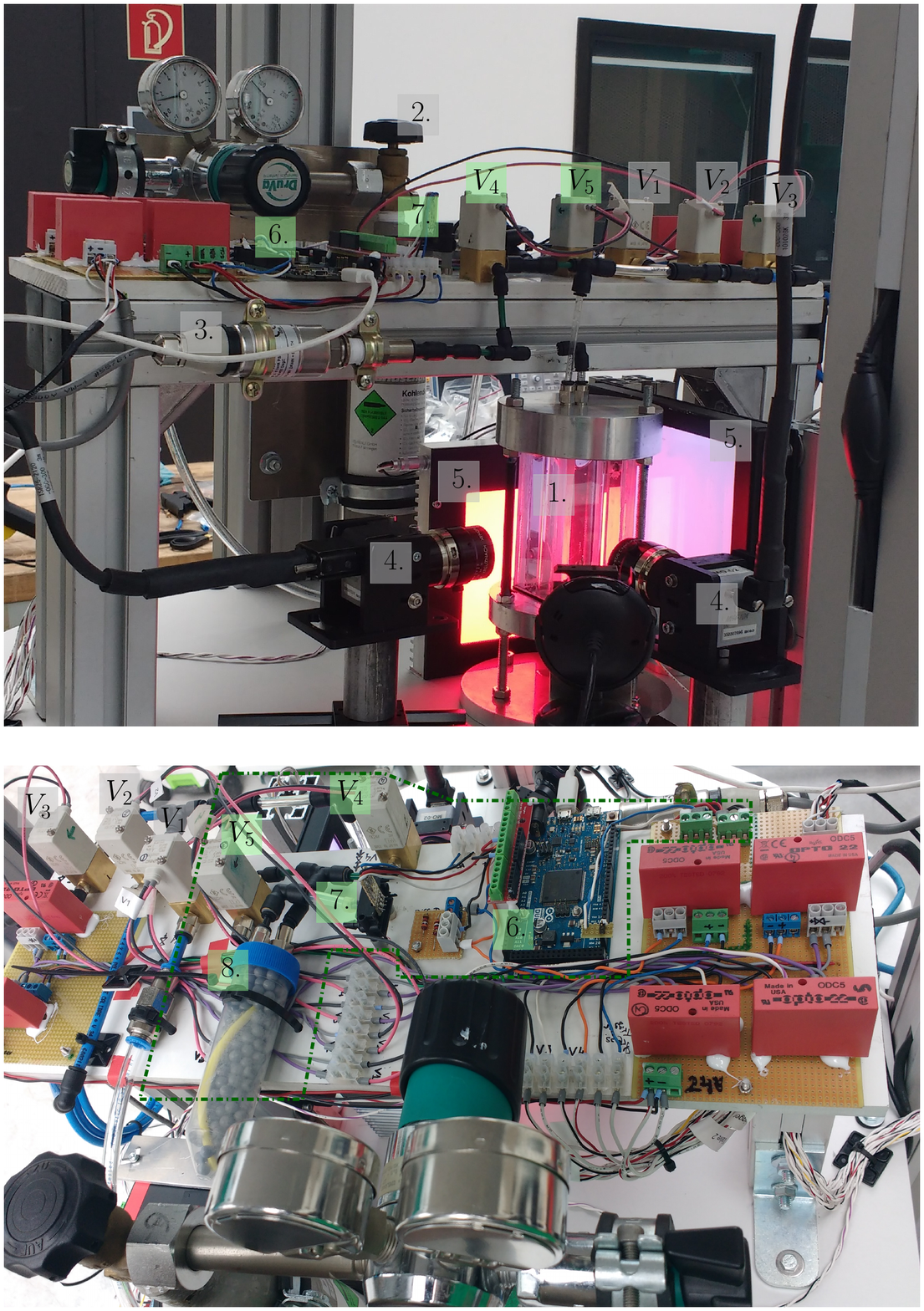}
\caption{\label{fig:photo}\textsl{Front and top view of the experiment inside the capsule: 1. Measurement tank, 2. CO$_2$ bottle, 3. Pressure sensor, 4. High-Speed Cameras (Photrom Fastcam MC-2), 5. LED screen, 6. Arduino board, 7. Differential Pressure Sensor, 8. Expansion tank. (The elements marked in green color correspond to the extended set-up which will be used in the future).}}
\end{figure}

The experiment consists in the formation of a bubble cloud in a supersaturated liquid by spark-induced cavitation and then, observing the development of the cloud using high-speed imaging in microgravity conditions. The layout of the experimental set-up is shown in figures \ref{fig:sketch} and \ref{fig:photo}. 

The measurement chamber, where the foam evolves, is the main component of the experiment. Its body is a cylinder of Pyrex glass (24.4 mm of diameter and 101.1 mm of height) and contains CO$_2$-supersaturated water. Around the cylinder there is a rectangular prism that is filled with degassed water. The purpose of this jacket is to avoid the optical aberration caused by a cylindrical container. At the top of the tank, there is a line which is connected to a pressurized CO$_2$ gas bottle through an electrovalve ($V_1$). In this line, there are two more electrovalves ($V_2$ and $V_3$) that connect the chamber to the ambient pressure to depressurize the chamber. Downstream of the electrovalve $V_2$ there is a reduction valve to achieve a smoother depressurization of the tank before the experiment starts. This is necessary since abrupt depressurization induces bubble growth at unwanted locations in the measurement tank. In addition, there is a pressure sensor (Gems, 220RAA6002F3DA, 0-6 bar) in order to control the pressure during the pressurization and depressurization of the tank. Near the bottom of the measurement tank, there are two copper filaments of 100 $\mu$m in diameter. These thin naked copper wires are used as electrodes, that are connected to the spark generator device that discharges a large capacitor in a very short time ($\sim$400 $\mu$s). The spark generator reproduces the discharge circuit described in Willert et al. \cite{circuit} but replacing the LED by the electrodes, as suggested by Goh et al. \cite{spark}. A capacitor (2200 $\mu$F) is charged through a power supply (30-35 V) and  is discharged thought a fast MOSFET power transistor when it receives a TTL trigger signal.
This discharge induces cavitation, and the collapse of the imploding bubble generates the bubble cloud which is the target of the experiment. 
The supersaturated liquid has been prepared in the installation designed by Enr\'{\i}quez et al. \cite{carbonatedwater} at the University of Twente. In this way, we can control the saturation level of the liquid. However, due to the manipulation during the filling of the measurement tank, the CO$_2$ concentration is lower than the initial concentration in the preparation.

\subsection{\textit{Experimental procedure}}
\label{procedure}
Before filling the measurement tank with carbonated water, we flush the chamber and the electrodes with alcohol to reduce bubble formation at the walls. The electrodes are in contact inside the tank and connected to the $Spark$ $Generator$ device. Initially, all the electrovalves are closed. Then, electrovalve $V_1$ is opened and the CO$_2$ gas fills up the measurement tank up to about 1.8-2 bar of pressure. The purpose is to dissolve all the bubbles that may have appeared in the chamber during its filling. After approximately 30-40 minutes, the electrovalve $V_1$ is closed. Now, electrovalve $V_2$ is opened, thus exposing the chamber at ambient pressure. Then, the capsule is dropped and, when it achieves microgravity conditions, the electrodes spark and the cameras are triggered. At this point, the experiment starts. 

\subsection{\textit{Extended setup}}
\label{Extension_setup}
\begin{figure}
\includegraphics[width = 1\columnwidth,trim= 100 0 100 0 ,clip]{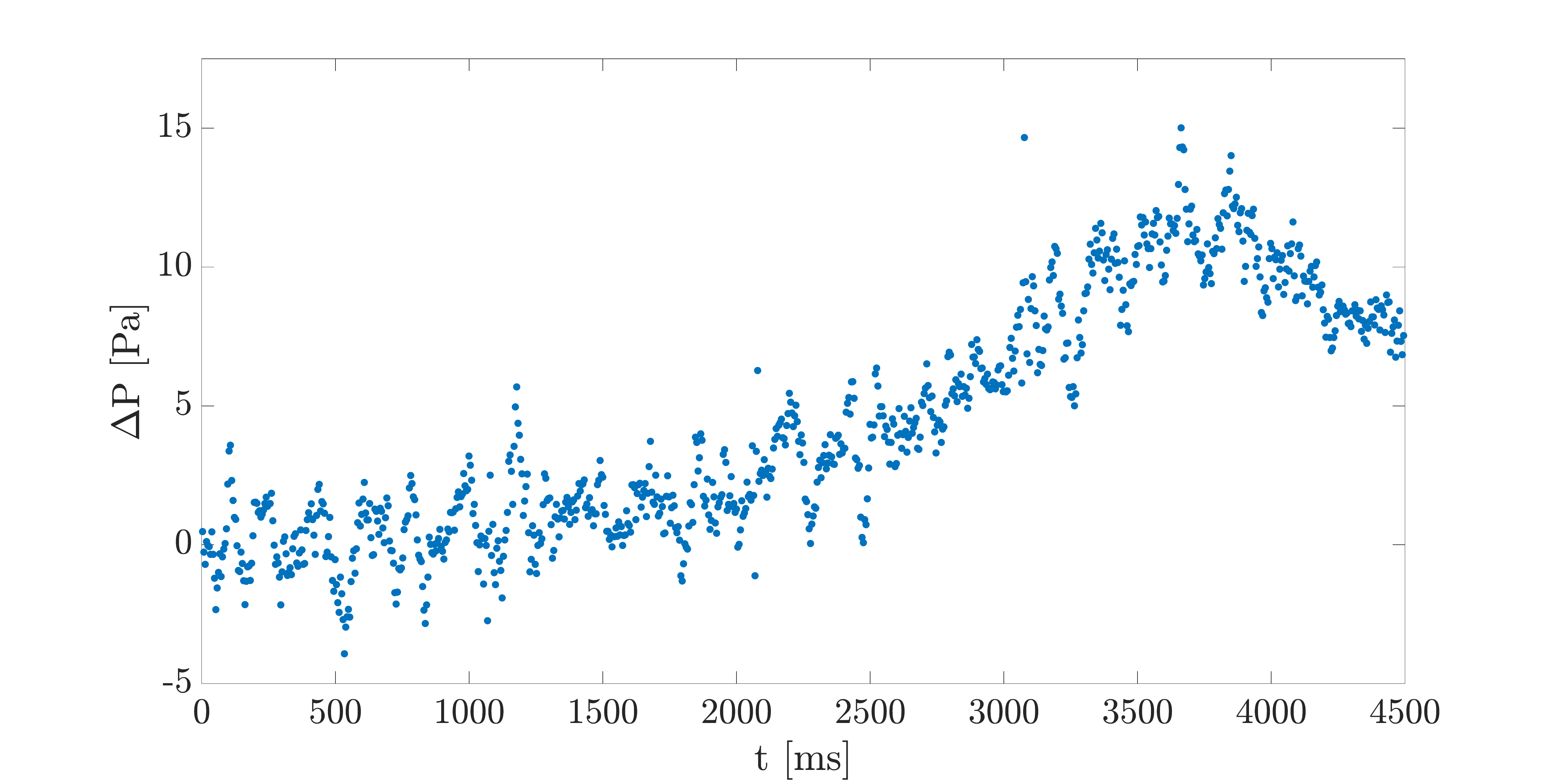}
\caption{\label{sensiron} \textsl{Example of pressure measurement taken with the Sensirion $SDP610 \pm25Pa$ differential pressure sensor. Mean data rate: 203 Hz. The measurement chamber is initially at ambient pressure and is connected to the atmosphere solely through the differential pressure sensor. At $t = 1$ s a 200 mJ pulse from a YaG laser is focused inside the chamber, which induces cavitation and the subsequent growth of the bubble cloud.}}
\end{figure}

In order to measure the time evolution of the total volume of exsolved gas in the measurement chamber, the following system has been designed: at the top of the measurement chamber, there is another line which connects through a capillary tube to a second, gas-filled vessel (expansion tank) in order to allow the liquid-gas mixture to freely expand during the experiment. As the bubble cloud grows inside the liquid, the free surface advances into the measurement line thus compressing the gas inside. It is easy to see that the overpressure satisfies $\left|\frac{\Delta P}{P_0} \right| \simeq \left|\frac{\Delta V}{V_0}\right|$, where $P_0$ is the initial pressure in the line (ambient), $\Delta P$ is measured by a differential pressure sensor (Sensirion, $SDP610 \pm25Pa$), $V_0$ is the initial gas volume in the line and $\Delta V$ is the volume of the exsolved gas. In this line, we place an expansion tank with a relatively large volume $V_0$ to act as a buffer, since the determination of the initial gas volume in the line is not feasible.
The differencial pressure sensor starts to measure a few milliseconds before the spark is triggered, and these measurements are stored in an Arduino Due board, which manages the sensor. In this line, there are two more electrovalves ($V_4$ and $V_5$) to control the pressure conditions in the system (see Fig. \ref{fig:photo}). As a consequence, the experimental procedure changes sligthly. When the electrovalve $V_2$ is opened, half a second after the $V_5$ is opened too, hence the pressure in the expansion tank will be at ambient pressure. At this time, the differential pressure sensor starts to measure and the following steps are the same as described in subsection \ref{procedure}. Figure \ref{sensiron} shows an example of pressure measurement acquired in an experiment in normal gravity conditions. The measurement chamber was initially at ambient pressure and connected to the atmosphere through a differential pressure sensor. Thus, after triggering cavitation with a laser pulse (occurring at $t = 1$ s), a bubble cloud is formed that grows due to diffusion and buoyancy-driven advection as explained in \cite{RodriguezRodriguez_etalPRL2014}. As the cloud volume increases, the liquid surface compresses the air inside the expansion tank as shown in Figure \ref{sensiron}. Interestingly, after $t \approx 3.5$ s the pressure decreases, coinciding this with the moment when the bubble cloud reaches the liquid free surface. The decrease of the pressure is due to the fact that the differential pressure sensor allows some gas to flow through it to the ambient, since it measures the pressure difference precisely based on this flow \cite{Homan_etalPRE2014}.

\section{Results and Discussion}

We report in this section preliminary results on the growth rate of the radius of the individual bubbles, the 3D reconstruction of the structure of the cloud and the time-evolution of the total gas volume in the cloud. The latter is obtained through the analysis of the mean grey level of the images, which will allow us to determine quantitatively the time evolution of the cloud's volume. The purpose of this section is to illustrate the kind of quantitative information that can be obtained with the experimental set-up described in this paper. The drops whose results are shown in Figs. \ref{fig:Radius}, \ref{fig:3D} and \ref{MGL_V} were carried out between June 27$^{\rm th}$ and July 1$^{\rm st}$, 2016.

\subsection{\textit{Radial expansion of individual bubbles}}

\begin{figure}
\centering
\includegraphics[width =1\columnwidth,trim= 20 0 500 0 ,clip]{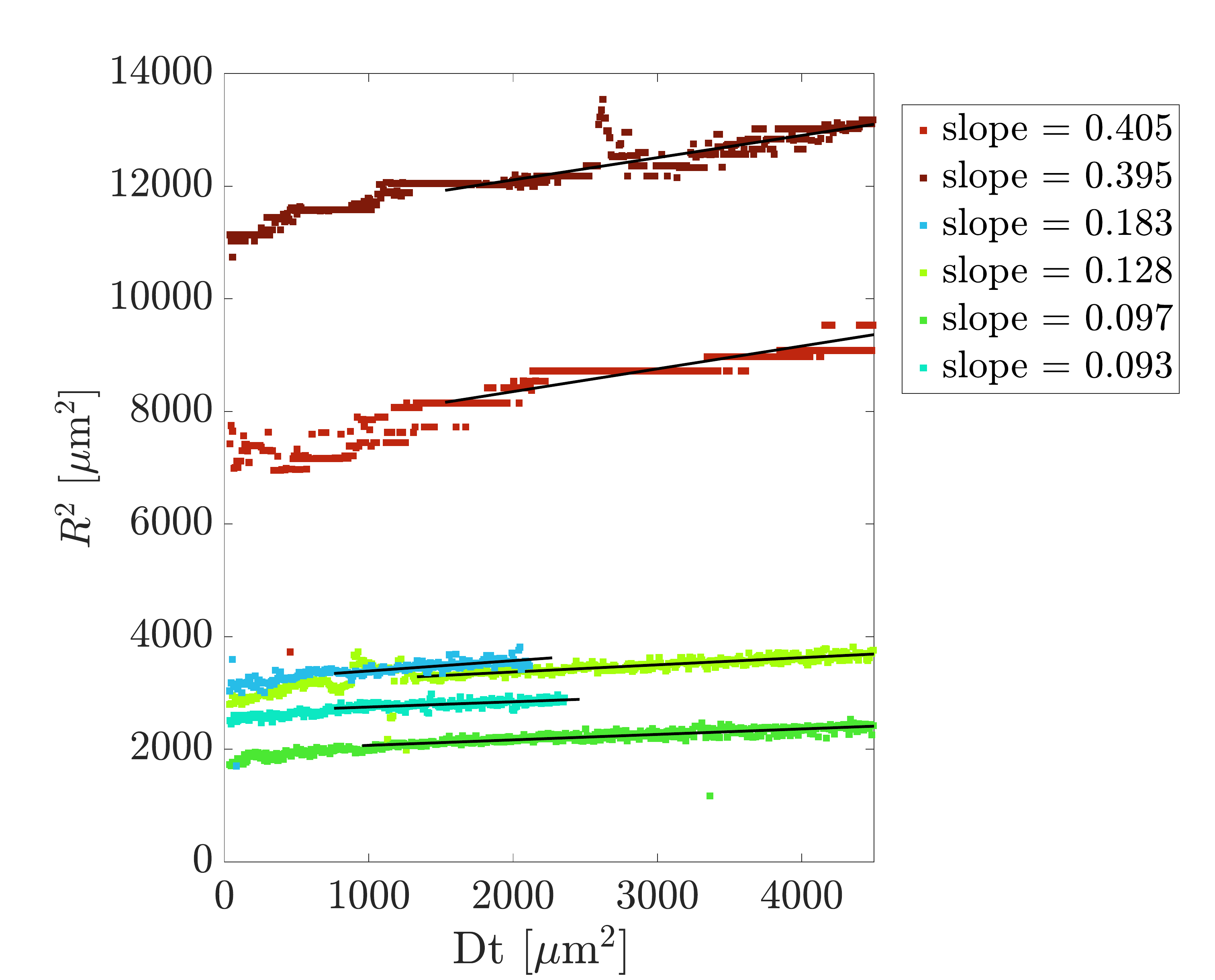}
\caption{\label{fig:Radius}\textsl{Time evolution of the squared radius of different bubbles in the experiment. The red square symbols represent the evolution of two bubbles attached to the electrodes as figure \ref{fig:comparation} shows in the pictures e - f, and which can be considered as isolated, while the bluish square symbols shows the evolution of several bubbles inside the cloud.}}
\end{figure}

We start from the well-known Epstein--Plesset equation \cite{3}, \cite{book}. It describes the growth rate of the bubble radius as a function of the properties of the gas and the level of saturation in the liquid. For our purpose, it is useful to rewrite it in terms of the square of the radius. Thus, after some mathematical manipulations we obtain:
\begin{equation}
\label{eq:R2}
\frac{1}{2} \frac{dR^{2}}{dt} = D \frac{\Delta C}{\rho_g} \left(1 + \frac{R}{\sqrt{\pi D t}} \right) ,
\end{equation}
where $D$ is the diffusivity of CO$_2$ in water, $\rho_g$ the density of the gas inside the bubble and $\Delta C$ the difference in concentration of CO$_2$ between the bubbles in the surface and the bulk fluid. Numerical integration of equation (\ref{eq:R2}) reveals that, for times of the order of $R_0^2/D$ or longer, with $R_0$ the initial bubble radius, the second term in the right hand side of equation (\ref{eq:R2}) approaches a constant \cite{3}, thus the square of the radius grows linearly with time: 

\begin{equation}
\label{eq:line}
R^{2} \sim F\left(\frac{\Delta C}{\rho_g}\right) Dt = F\left(\Lambda (\zeta -1)\right) Dt,
\end{equation}
where $\Lambda = K_h R_g T_\infty$, $K_h$ is Henry's constant, $R_g$ the gas constant and $T_\infty$ the temperature. Physically speaking, $\Lambda$ is a constant which measures the gas solubility. In the parenthesis, $\zeta$ is the supersaturation level which measures the amount of dissolved CO$_2$ available for bubble growth.  The function $F(x)$ is given by:
\begin{equation}
F(x) = \frac{x}{\sqrt{\pi}} + \sqrt{\frac{x^2}{\pi} + 2x}.
\end{equation}

Note that, in these calculations, the effect of surface tension on the gas pressure has been neglected, as bubbles are much larger than $2\sigma/P \approx 0.8$ $\mu$m, the size at which capillary overpressure, $2\sigma/R_0$, becomes equal to the ambient one, $P$. Indeed, taking $\sigma = 0.0434$ N/m, the Laplace overpressure is at most about 2\% of the ambient one even for the smallest bubbles reported here.

To measure the growth of the bubble radius in the experiments, we track individual bubbles in the cloud using custom-made image processing software implemented in Matlab. The time evolution of the bubble radii for different bubbles in the same experiment is shown in figure \ref{fig:Radius}. In some experiments, bubbles are attached to the electrodes. We consider these bubbles to be isolated as they are far away from others whereas the area in touch with the electrode is small. Consequently, their growth rate gives us an estimation of the saturation level (Fig. \ref{fig:Radius}).
Indeed, the growth rates of the squared radius of the bubbles considered as isolated are around three times larger than those of bubbles found inside the cloud. Moreover, the growth rate of the bubbles on the electrodes are nearly the same, whereas in contrast, growth rates for the bubbles inside the cloud differ in a visible way (Fig. \ref{fig:Radius}). This variability of growth rates is consistent with the fact that these bubbles compete for the available CO$_2$ in their surroundings and this competition for the CO$_2$ provides information about how an individual bubble interacts with the rest of the bubbles in the cloud.\\

\begin{figure}
\centering
\begin{subfigure}[b]{1\columnwidth}
\includegraphics[width =1\columnwidth,trim = 225 0 225 20,clip]{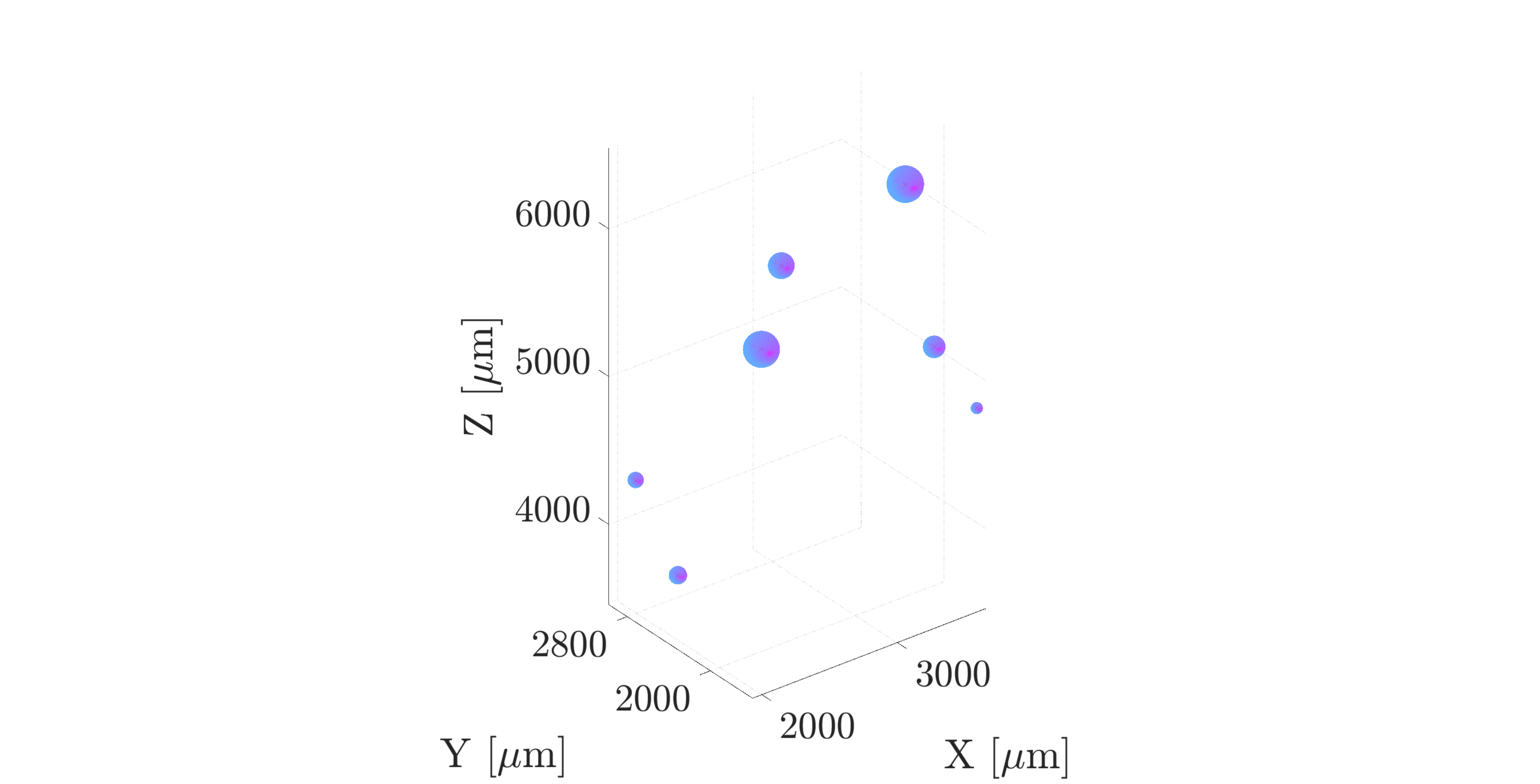}
\end{subfigure}
\caption{\label{fig:3D} \textsl { Using digital image processing, we detect the position of large bubbles at each instant of time in the images of the two cameras. This allows us to detect reconstruct the 3D structure of the bubble cloud, or at least of those bubbles large enough to be tracked.}}
\end{figure}

In order to get quantitative data to validate future models of bubble cloud growth, we take advantage of the fact that the experiment uses two high-speed cameras forming a 90$^{\rm o}$ angle to reconstruct the 3D structure of the cloud (see Fig. \ref{fig:3D} for an example). This information will allow us to relate the local level of saturation with the position of the bubble in the cloud.

\subsection{\textit{Determining the gas volume of the cloud}}

Although tracking individual bubbles yields very useful quantitative information, this technique can only be applied to relatively large bubbles. Nonetheless, a significant volume of the cloud corresponds to very fine and small bubbles that lie below the spatial resolution of our high-speed cameras.
As an alternative way to determine the time evolution of the total gas volume of the cloud, we analyze the mean grey level of the images. 
This mean grey level, $MGL$, can be defined as
\begin{equation}
MGL = \frac{\sum\limits_{i=1}^H\sum\limits_{j=1}^W Im_{i, j}}{H  W},
\end{equation}
where $H$ and $W$ are the height and width of the image matrix, $Im$, respectively. Next, we must  establish a relation between this mean grey level and the volume, in other words, we must obtain a calibration curve. This can be done by generating a bubble cloud with a well-controlled volume via electrolysis. 

\subsubsection{\textit{Calibration}}

Indeed, using Faraday's law for electrolysis \cite{Faraday_law,Faraday_law1}, we are able to predict the volume of the gas generated provided the current flowing through the electrodes is known.
The expression of the Faraday's law for electrolysis is
\begin{equation}
m = \left(\frac{Q}{F}\right) \left(\frac{M}{z}\right),
\end{equation}
where $m$ is the mass of the substance liberated at an electrode, $Q$ is total electric charge that has flowed through the electrolyte (the liquid), and that can be obtained as the time integral of the current; $F = 96500$ C$\cdot$mol$^{-1}$ is Faraday's constant, $M$ is the molar mass of the gas (Hydrogen in our case) and $z$ is the number of electrons transferred per ion. The mass, $m$, divided by the molar mass, $M$, is the number of moles, $n$.
In order to obtain the volume, $V$, we use the ideal-gas law $P V = n \bar{R} T_\infty$ where, $P$ is the pressure of the gas (nearly ambient here), $\bar{R}$ is the universal gas constant and $T_\infty$ is the temperature.

Thus, we carried out calibration experiments producing bubble clouds by electrolysis in the same experimental chamber used for the drops. The procedure was the following: the measurement chamber is filled up with clean water and the electrodes are separated and connected to a current source. This produces a known volume of gas which is then filmed with the high-speed camera. In order to measure the current in the circuit, a resistor ($\Omega_c = 18 \Omega$) is placed in series with the electrodes. The voltage across the resistor is measured with an oscilloscope. We use deioned water to which a small amount (15 grams per liter) of potassium carbonate was added in order to make it conductive \cite{PhD_Ana}.
The overall reaction of the electrolysis of the water is \cite{waterelectr},
\begin{equation}
2 H_{3} O^{+} + 2e^{-} \longrightarrow  H_{2} (g) + 2 H_{2}O, 
\end{equation}
so, $z$ will be equal to 2.

\begin{figure}
\centering
\includegraphics[width=1\columnwidth,trim = 60 0 30 0,clip]{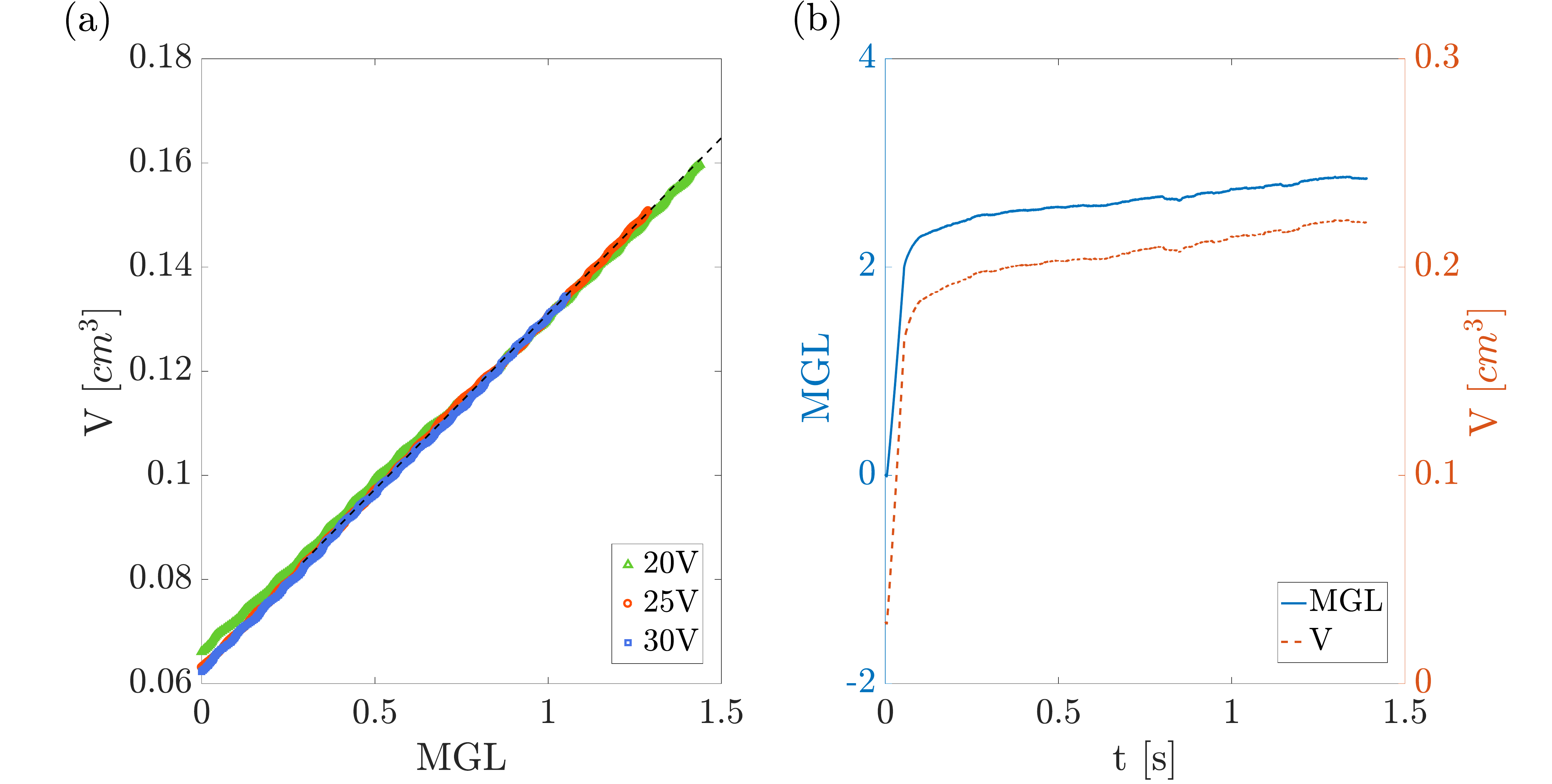}
\caption{ \label{MGL_V} \textsl {Plot (a) represents the relation between the mean grey level and the gas volume generated by electrolysis at different voltages, namely 20 $V$, 25 $V$ and 30 $V$, P = 1 $atm$ and T = 288 $K$. The dashed line shows the callibration curve. In plot (b), the blue solid line presents the time evolution of the mean grey level of one experiment and the orange dashed line shows the time evolution of the gas volume calculated according to the callibration curve. In both plots, the data has been filtered using a moving average filter to remove high-frequency noise.} }
\end{figure}

We show the mean grey level of the images with the volume produced by electrolysis at different voltages (see Fig. \ref{MGL_V}a). The mean grey level increases linearly with the volume. Moreover, the curves for the different voltages and different experimental realizations (3 per voltage) overlap, which proves the reliability of the results. Indeed, although the bubble size distributions show some variation for the different voltages, these changes do not affect the calibration curve. It should be pointed out that, in the calibration curves (Fig. \ref{MGL_V}a) the first tens of milliseconds upon starting electrolysis have been excluded, since at those early times the Hydrogen has remained in dissolution and does not form bubbles.

Therefore, making use of the calibration curve obtained, the time evolution bubble cloud volume evolution in the drops can be estimated with the analysis of the mean grey level. As an example, figure \ref{MGL_V}b shows the evolution of the mean grey level and the volume for drop $N^{\circ}$4 of the campaign.


\section{Conclusion}

The goal of this experiment is the study of the diffusion-driven growth of a bubble cloud in a CO$_2$ supersaturated water solution at times much longer than a few hundreds of milliseconds, when gravity becomes dominant in normal conditions on Earth. In our preliminary experiments under microgravity conditions, the evolution of the cloud can be observed for more than 1 second (Fig. \ref{fig:comparation}). Still the cloud moves in the microgravity tests as a consequence of the residual velocity resulting from the implosion. Nonetheless, although the typical Peclet number, $Pe = V R_0 / D$, computed with the measured bubble velocity, $V$, for the bubbles tracked in this study is relatively large ($Pe \approx 60$), the linear relation observed between the square of the bubble radii, $R^2$ and the time $t$ suggests that advection plays a small role in bubble growth. This is consistent with the fact that, although the bubble cloud may translate as a whole, the relative velocity between each bubble and the fluid in its vicinity is much smaller than $V$. Consequently, we can apply the Epstein--Plesset equation \cite{3} to predict bubble growth. Interestingly, the different slopes of the $R^2$ vs. $t$ curves allow us to estimate the local concentration of CO$_2$ that every bubble experiences, which can later be connected to its location inside the cloud thanks to the 3D-reconstruction.

Furthermore, the analysis of the grey level can be used to estimate quantitatively the time evolution of the total volume of gas in the cloud. This has been checked by calibrating the mean grey level using images where the gas volume was generated by electrolysis, so it could be accurately determined at all times.

In summary, the experiment described here will allow us in future drop tower campaigns to gather very relevant quantitative information on the diffusion-driven growth of a cloud of bubbles in a gas-supersaturated liquid solution.


\begin{acknowledgements}
The authors thank the team from the ZARM Drop Tower Operation and Service Company (ZARM FAB mbH) for valuable technical support during the finalization of the setup and the measurement campaign. The European Space Agency is acknowledged for providing access to the drop tower through grant HSO/US/2015-29/AO "Diffusion-driven growth of a dense bubble cloud in supersaturated liquids under microgravity conditions". This work was supported by the Netherlands Center for Multiscale Catalytic Energy Conversion (MCEC), an NWO Gravitation programme funded by the Ministry of Education, Culture and Science of the government of the Netherlands. Finally, we wish to thank the Spanish Ministry of Economy and Competitiveness for supporting the building of the experimental facility through grants DPI2014-59292-C3-1-P and DPI2015-71901-REDT, partly funded through European Funds.
\end{acknowledgements}

%


\end{document}